\documentclass[amssymb,amsmath,aps,showpacs,twocolumn,floatfix,nofootinbib,showpacs]{revtex4}

\usepackage{graphicx}
\usepackage{color}
\usepackage{soul}
\usepackage{latexsym}

\newcommand{\lsim}{\lesssim}
\newcommand{\gsim}{\gtrsim}

\def\lsim{\mathrel{\raise.3ex\hbox{$<$\kern-.75em\lower1ex\hbox{$\sim$}}}}
\def\gsim{\mathrel{\raise.3ex\hbox{$>$\kern-.75em\lower1ex\hbox{$\sim$}}}}

\def\beq{\begin{equation}}
\def\eeq{\end{equation}}
\def\beqn{\begin{eqnarray}}
\def\eeqn{\end{eqnarray}}
\def\bea{\begin{eqnarray}}
\def\eea{\end{eqnarray}}
\def\be{\begin{equation}}
\def\ee{\end{equation}}
\newcommand{\fslash}[1]{{#1 \kern -0.7em/ \kern 0.1em}}

\begin{document}

\voffset 1.25cm

\title{Detecting light leptophilic gauge boson at BESIII detector}

\author{Peng-fei Yin, Jia Liu and
Shou-hua Zhu}

\affiliation{Institute of Theoretical Physics \& State Key
Laboratory of Nuclear Physics and Technology, Peking University,
Beijing 100871, China}

\date{\today}

\begin{abstract}
The $ O(GeV)$ extra $ U(1)$ gauge boson named U-boson, has been
proposed to mediate the interaction among leptons and dark matter
(DM), in order to account for the observations by PAMELA and ATIC.
In such kind of models, the extra $U(1)$ gauge group can be chosen
as $U(1)_{L_i-L_j}$ with $L_i$ the $i-$th generation lepton number.
This anomaly-free model provides appropriate dark matter relic
density and boost factor required by experiments. In this work the
observability of such kind of U-boson at BESIII detector is
investigated through the processes $ e^ + e^ - \to U\gamma$,
followed by $U\rightarrow e^+e^-$, $U\rightarrow \mu^+\mu^-$ and
$U\rightarrow \nu\overline{\nu}$. In the invisible channel where
U-boson decays into neutrinos, BESIII can measure the coupling of
the extra $ U(1)$ down to $ O(10^{ - 4} ) \sim O(10^{ - 5} )$
because of the low Standard Model backgrounds. In the visible
channel where U-boson decays into charged lepton pair, BESIII can
only measure the coupling down to $ O(10^{ - 3} ) \sim O(10^{ - 4}
)$ due to the large irreducible QED backgrounds.

\end{abstract}



\maketitle

\section{Introduction}

Over the past several years, the existence of dark matter (DM) has
been confirmed by many astronomical observations, but its exact
nature is still unknown. The annihilation or decay products of DM
like photons, neutrinos and antimatter particles may be observed by
DM indirect detecting experiments. Among these methods, detecting
positrons or anti-protons from DM is a challenge due to the
background induced by cosmic-rays or other astrophysical sources and
various uncertainties during anti-matter propagation. Recently
PAMELA satellite experiment reported an excess in flux ratio of
positrons to the sum of electrons and positrons around $10 GeV$ to
$100 GeV$ while the flux ratio of anti-proton to proton has no
obvious deviation from the prediction from cosmic-rays
\cite{Adriani:2008zq}. In addition, the ATIC reported the total flux
of electrons plus positrons spectrum measurement up to $1 TeV$ in
which there is a bump over the background around $300 GeV$ to $800
GeV$ \cite{:2008zz}. These results provide a new perspective to DM
research which has quite different features compared to "popular"
candidates in the literatures.

The ATIC electron/positron excess suggests a heavy dark matter
around $O(1) TeV$. To accommodate the PAMELA results, the DM seems
to be leptophilic to avoid the anti-proton excess (If the measured
anti-matter particles are produced in the nearby DM subhalo
\cite{Hooper:2008kv} or we use some special cosmic-ray propagation
models \cite{Grajek:2008jb}, this constraint may be loosen). For the
annihilating DM, there is a significant mismatch, namely the
expected thermal-DM annihilation cross-section is much smaller than
those required by PAMELA and ATIC measurements. There is a class of
DM scenarios which can satisfy all past experiments. The
extraordinary prediction of this scenario is that  there are some
new light scalars or gauge bosons to mediate DM sector
(eg.\cite{ArkaniHamed:2008qn,Pospelov:2008jd}). The exchange of
light mediator should increase the DM annihilation cross section at
low velocity, such as in the Galaxy today, comparing with the
velocity in the epoch of freeze-out due to the so-called "Sommerfeld
enhancement" \cite{ArkaniHamed:2008qn,Pospelov:2008jd,sommerfeld}.
Moreover if such mediators are the only products from DM
annihilation and they are light enough to forbid the decays into
baryons, the DM will produce only charged leptons. In this scenario
such mediator may actually interact with all the Standard Model(SM)
particles through the mixing with SM $U(1)_Y$ gauge field or Higgs
field
(eg.\cite{ArkaniHamed:2008qn,Pospelov:2008jd,Chun:2008by,Baumgart:2009tn}).
Instead we can impose the new symmetry to make sure that the
mediator only interacts with leptons, at least at the tree-level
\cite{Fox:2008kb,Baek:2008nz}. For example the mediator can be the
gauge boson of an extra $U(1)_{L_i-L_j}$ \cite{He:1990pn,Bi:2009uj},
where $L_i$ is the number of i-th generation of lepton. This model
is anomaly free due to the cancelation between two generation of
leptons with opposite $U(1)$ charge. In this paper we will focus on
the search on this kind of light new gauge boson at BESIII detector.

It is the well-motivated scientific goal to search for such light
boson $X$ at the low-energy $e^+e^-$ colliders due to its possible
leptophillic feature. Obviously $X$ should not contradict with the
known measurements, such as anomalous magnetic moments of charged
leptons $g-2$, $\nu-e$ scattering cross section,
$etc$.\cite{Fayet:2007ua}. Provided that the X is light, the
interactions between $X$ and SM particles should be weak. However
the signals of X at colliders may be heavily polluted by large QED
backgrounds. On the other hand, the invisible decay of X, i.e. X
decays into final states which do not interact with detector, is
promising because the irreducible SM backgrounds arise from neutrino
which is suppressed by $O(Q^2/m_Z^2)$ for low energy linear collider
\cite{Zhu:2007zt}. It should be emphasized that missing energy
measurements are always challenging from the experimental point of
view. Thus the detection of light gauge boson at the low-energy
experiments is a great challenge. As a result, large luminosity is
required in order to collect enough events and suppress QED
backgrounds.

In this paper, we extend our previous investigation on light new
gauge boson \cite{Zhu:2007zt} to $O(GeV)$ at BESIII detector. In the
previous work \cite{Zhu:2007zt}, the possibilities of detecting
$O(MeV)$ new gauge boson, usually called U-boson in the literature,
has been scrutinized. Such $O(MeV)$ U-boson is used to explain the
excess of 511 keV photon line which was observed by INTEGRAL
\cite{Boehm:2003bt}. In fact, research on extra light gauge boson
has a long history
\cite{EarlyUBoson,Gninenko:1998pm,Borodatchenkova:2005ct,Bouchiat:2004sp,Zhu:2007zt,Fayet:2007ua,Chen:2007uv}.
The main production process at low-energy colliders can be
$e^+e^-\rightarrow \gamma U$. If the U-boson have invisible decay
channel to DM \cite{Boehm:2003hm}, the detection will benefit from
small SM backgrounds $e^+e^-\rightarrow \gamma \nu\overline{\nu}$.
If the U-boson decays into charged lepton pairs, the SM background
is $e^+e^-\rightarrow \gamma l^+l^-$. For the SM backgrounds,
$m_{l\bar l}$ is smoothly distributed. Instead the signal peaks
around the mass of U-boson, the invariant mass of lepton pair should
be utilized to distinguish the events from the backgrounds. Thus the
detection of such U-boson strongly depends on the mass resolution
and the integrated luminosity of the experiment. Provided that the
U-boson obtained the mass via the spontaneously symmetry breaking of
a new scalar field, there is at least one extra scalar particle in
the particle spectrum. Such scalar particle may also be detected at
the low energy colliders, provided its mass are within the reach of
these colliders. In this paper we do not focus on this investigation
and allocate it to the further studies.

Recently, some authors did the investigations of such new light
gauge boson at low energy experiments
\cite{ArkaniHamed:2008qp,Pospelov:2008zw,Batell:2009yf,Essig:2009nc,Reece:2009un}.
The discussions at $e^+e^-$ colliders often concentrate at B-factory
and $\phi$-factory. They investigated the processes of
$e^+e^-\rightarrow \gamma U$ and meson decay
\cite{Pospelov:2008zw,Reece:2009un}, scalar strahlung
\cite{Batell:2009yf} $etc$. in the context of U-boson mixing with SM
particle. In this paper, we consider a gauge boson in the model with
extra $U(1)_{L_i-L_j}$ gauge group which will be described in the
next section.

This paper is organized as following. In the section II, we describe
the $U(1)_{L_i-L_j}$ model with an extra light U-boson, as well as
the constraints from current low energy experiments. In Section III,
we discuss the influences of such gauge boson in the epoch of DM
freeze-out and in the Galaxy today. We found that the coupling
between DM and U-boson is about $O(10^{-1})$ while the allowed
couplings between SM fermion and U-boson are small. In Section IV,
we simulated the signals and backgrounds for U-boson at the BESIII
detector. The conclusions and discussions are given in the last
section.

\section{The model}

We adopt the models in the Ref. \cite{Fox:2008kb,He:1990pn} in which
the extra U(1) charge can be $ L_e - L_\mu$, $ L_e - L_\tau$ or $
L_\mu - L_\tau$. This gauge group is broken by a SM singlet scalar
Higgs field $S$ which gives the U-boson mass around $O(GeV)$. We
define $g_A=g''\cdot C_A$ ($A$ denotes any particle), where $g''$ is
gauge coupling constant of the extra $U(1)$ interaction, and $C_A$
is the extra U(1) charge of a particle. Besides the SM Lagrangian,
the additional Lagrangian may be written as
\begin{eqnarray}
\mathcal{L} &=&  - \frac{1}{4}F_{\mu \nu}^{\prime \; 2}  + \frac{\kappa }{2}F_{\mu \nu}^{\prime} F^{\mu \nu} + \sum\limits_l {\bar l} (i\fslash{D} - m_\psi  )l \nonumber\\
&+& |D_\mu S|^2 - V(S) + \lambda _{SH} (S^\dag  S)(H^\dag  H)+
\mathcal{L}_{DM} ,
\end{eqnarray}
where $F_{\mu \nu}^{\prime}$ and $F_{\mu \nu}$ are the field
strength for U-boson and the gauge boson B of $ U(1)_Y$,
respectively. $D_\mu=\partial_\mu + ig_A U_\mu$ is the covariant
derivative, $V(S)$ is the $S$ field potential, and
$\mathcal{L}_{DM}$ is the Lagrangian which describes the
interactions of DM sector.

Generally speaking, the  mixing parameters $\kappa$ and $\lambda
_{SH}$ among the new bosons and SM bosons are not zero, but they
tend to be small due to the limits from low energy experiments. If
we forbid these mixing parameters at the tree level, they will be
induced from the higher order contributions. But the interactions by
higher order contributions are usually small  \cite{Fox:2008kb}. In
this work, we neglect the mixing effects for simplicity. The scalar
potential can be written as $\mu_{S}^{2} |S|^2+ \lambda_S |S|^4$.
After the spontaneous symmetry breaking of $S$ with the vacuum
expectation value $v_S/\sqrt{2}$, U boson obtains mass to be $ m_U =
g''v_S/2$. In our work we do not work on the possibility of
searching light scalar $S$ which has been discussed in the Ref.
\cite{Batell:2009yf}. There is also an extra heavy particle $\chi$
as the candidate of DM in the model. The particle $\chi$ can be a
scalar or vector-like fermion (such choice is the simplest way to
construct anomaly-free model) with mass around $1TeV$ which is
favored by ATIC experiment. The Lagrangian of DM can be written as
\cite{Fox:2008kb}
\begin{eqnarray}
\mathcal{L}_{DM}&=&
\left\{\begin{array}{cc}\;\;\;\; \bar{\chi}(i\fslash{D} - m_\chi)\chi \;\;\;\;\; \chi\; is\; fermion\\
|D_\mu \chi|^2-m_\chi^2|\chi|^2 \;\;\; \chi\; is\; scalar.
\end{array}\right.
\end{eqnarray}

The parameters $m_\chi$ and $g_\chi$ are important in the DM sector,
but they have negligible effects on the low-energy experiments. The
parameter $m_U$ is important in both sectors and it mediates the
interactions among the DM and the SM particles. Note that the
interactions between U-boson and leptons are vector-like and the
U-boson couples also with the neutrinos. In addition, the
$U-\ell-\bar\ell$ couplings $g_l $ are universal for two generations
of leptons. These couplings have been constrained by many known low
energy measurements. From these observations, the constraint on the
contributions to the anomalous magnetic moments of the charged
leptons $a_l=(g_l-2)/2$ induced by U-boson are very stringent. The
additional contributions from U-boson for a vector-like interactions
is given by \cite{Fayet:2007ua}
\begin{equation}
\delta a_l^V\simeq \frac{g_l^2}{4\pi^2}\int_0^1dx \frac{m_l^2
x^2(1-x)}{m_l^2 x^2+m_U^2(1-x)}.
\end{equation}

For the $g_e$, following the discussion in Ref. \cite{Fayet:2007ua,
Hanneke:2008tm}, we have the constraint $\delta
a_e^V<1.5\times10^{-11}$. For the $g_\mu$ and $g_\tau$, we impose
conservative constraints as $\delta a_\mu^V<2.6\times10^{-9}$ and
$\delta a_\tau^V<1.3\times10^{-2}$ by using the results from Ref.
\cite{Bennett:2006fi} and Ref.\cite{Abdallah:2003xd} respectively.
Another stringent constraint for vector-like coupling arises from
low-$|q^2|$ neutrino-electron scattering \cite{Allen:1992qe} as
$|f_\nu f_e|/m_U^2 < G_F$ \cite{Fayet:2007ua}. The universal
leptonic couplings can be written as $|f_e|^2/m_U^2 < G_F$. Combing
all these constraints, the coupling $g_l$ should be smaller than
$O(10^{-3})$.

\section{Light gauge boson interacting with DM}

From section II, we can see that the couplings among U-boson and SM
particles are small. However the couplings among U-boson and DM can
be large. It is quite natural to assume the main products of DM
annihilations are light U-bosons. In this section we will
investigate the magnitude of couplings among U-boson and DM from
cosmological point of view.

At the high temperature, the U-boson can reach the equilibrium with
the charged leptons and neutrinos. The thermal relic density of DM,
only for s-wave, is given by \cite{Jungman:1995df}

\begin{equation}
\Omega_\chi h^2\simeq \frac{1.07\times10^{9}}{g_\ast^{1/2}
m_{pl}(<\sigma v>/x_f)}, \label{relic}
\end{equation}
where $m_{pl}$ is the Planck mass of $1.22\times 10^{19} GeV$. The
$g_\ast$ is the effective number of relativistic degrees of freedom,
which is around 100 at the epoch when DM is freeze-out. The $<\sigma
v>$ is the thermally averaged DM annihilation cross section from
s-wave in unit of $GeV^{-2}$. The $x_f$ is related to the freeze-out
temperature $T_f$ and is defined as $x_f=m_\chi/T_f$, which can be
expressed as

\begin{equation}
x_f \simeq ln \frac{0.0955\; m_{pl}\; m_\chi\;<\sigma
v>}{\sqrt{g_\ast x_f}}.
\end{equation}

If the DM is fermion, the thermal averaged annihilation cross
section at freeze-out epoch is

\begin{equation}
\sigma v\simeq\frac{g_\chi^4}{16\pi m_\chi^2}
\frac{(1-\frac{m_U^2}{m_\chi^2})^{\frac{3}{2}}}{(1-\frac{m_U^2}{2m_\chi^2})^2}.
\end{equation}

It is obvious that the cross section only depends on $g_\chi$ and
$m_\chi$ when $m_U/m_\chi$ approches 0 for light U-boson here. For
the scalar DM case, the results can be written as

\begin{equation}
\sigma v\simeq\frac{g_\chi^4}{8\pi m_\chi^2}
(1-\frac{m_U^2}{m_\chi^2})^{\frac{1}{2}}.
\end{equation}

If the DM also carries extra $U(1)$ charge, the cross section
requires an extra factor $1/2$ for averaging initial DM charge
\cite{Pospelov:2008jd}. If the DM mass is $O(TeV)$, the correct
relic density requires $g_\chi \sim O(10^{-1})$. Comparing with
$g_l\leq O(10^{-3})$ depicted in last section, we can conclude that
DM should have much larger $U(1)$ charge than those of the SM
leptons (some possibilities to explain this feature have been
discussed in the Ref. \cite{Fox:2008kb}).

In the Fig.\ref{relic}, we show the possible parameter region which
satisfies the relic density $0.085<\Omega h^2<0.119$. From the
figure we can see that the $g_\chi$ is indeed $O(10^{-1})$ for DM
with mass of $O(TeV)$.

\begin{figure}[h]
\vspace*{-.01in}
\centerline{\includegraphics[width=3.3in,angle=0]{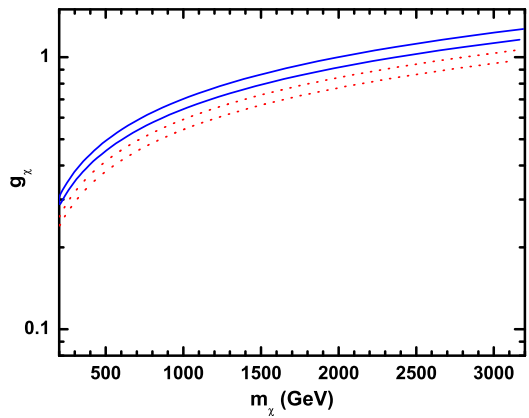}}
\vspace*{-.03in} \caption{The ellipses indicate the region of the
$m_\chi,g_\chi$ plane which satisfied the relic density
$0.085<\Omega h^2<0.119$, with $ m_U  = 0.5GeV$. Solid lines denote
for fermion DM and dash lines denote for scalar DM. \vspace*{-.1in}}
\label{relic}
\end{figure}

The light U-boson may enhance the DM annihilation cross section at
low velocity in the Galaxy today due to the non-perturbative effect
named "Sommerfeld enhancement"(a complete analysis can be found in
the Ref.\cite{ArkaniHamed:2008qn,Pieri:2009zi}). This
non-relativistic quantum effect arises because the two particle
wave functions are distorted away from plane wave by the presence of
a potential if their kinetic energy are low enough. In the language
of quantum field theory, it corresponds to the contribution of
ladder diagrams due to the exchange of some light scalars or gauge
bosons during two incoming DM particles undergoing some annihilation
reaction. In the non-relativistic limit, the exchange of a scalar
boson or a vector boson would give the same
result\cite{Iengo:2009ni}. This enhancement can be described by a
factor $S$ which is defined as a factor to multiply with the tree
level DM annihilation cross section, $ \sigma = \sigma _0 S$. This
factor is essential to interpret the difference between the DM cross
section required by the correct thermal relic density and the
PAMELA/ATIC positron anomaly.

In order to calculate S, we use the simplified quantum mechanical
method in literatures by solving the $ l = 0$ Schr\"{o}dinger
equation with an attractive Yukawa potential $V(r)=-\frac{\alpha}{r}
e^{- m_{U}^{\;} r }$ \cite{ArkaniHamed:2008qn,Pieri:2009zi},

\begin{equation}
\frac{1}{m_\chi}\psi''(r)+\frac{\alpha}{r}e^{-m_{U}^{\;}
r}\psi(r)=-m_\chi \beta^2 \psi(r) \label{radial},
\end{equation}
where $\psi(r)$ is the reduced two-body wave function, $m_\chi$ and
$m_U$ are the masses of DM and the light gauge boson respectively, $
\beta  = v/c$
 is the velocity of DM in the center-of-mass
frame and $ \alpha  = g_\chi ^2 /(4\pi ) $. The boundary condition
can be chosen as $ \psi '(\infty )/\psi (\infty ) = im_\chi  \beta$.
Then the Sommerfeld factor S is given by $ S = \left| {\psi (\infty
)/\psi (0)} \right|^2$. The behavior of $S$ depends on four
parameters $m_U^{\;}$, $m_\chi$, $\alpha$ and the velocity of DM
$\beta$.

Since the DM particles do not have monochromatic relative velocity,
a more realistic result needs to consider the speed distribution of
DM in the Galaxy. Here we assume the DM velocity distribution in the
halo as a single truncated Maxwell-Boltzmann distribution
$f(v)\propto v^2 exp(-\frac{v^2}{2\sigma_v^2}) $ with velocity
dispersion $\sigma_v$\cite{Bovy:2009zs}. The average of the
Sommerfeld enhancement over the distribution of relative velocities
in the halo is,

\begin{equation}
\overline{S}
=\sqrt{\frac{2}{\pi}}\frac{1}{\sigma_v^3}\int_0^{v_{esc}} dv\;v^2
exp(-\frac{v^2}{2\sigma_v^2})S(v).
\end{equation}

\begin{figure}[h]
\vspace*{-.01in} \centerline{
\includegraphics[width=3.3in,angle=0]{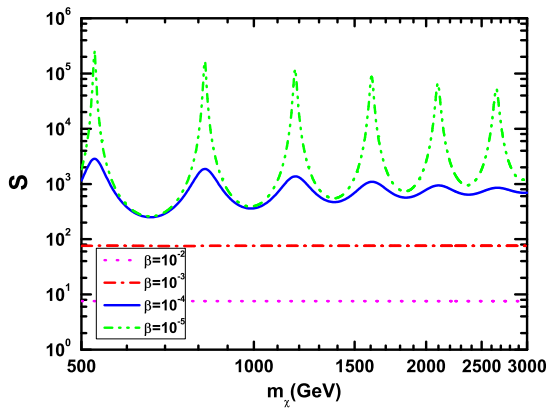}}
\vspace*{-.03in} \caption{The Sommerfeld enhancement factor S
 as a function of DM
mass. Here we choose $m_U=0.5GeV$ and $g_\chi=0.55$ ($ \alpha  =
2.41 \times 10^{ - 2}$ ). Four curves denote different DM velocity $
\beta$ as $10^{-2}$,
 $10^{-3}$,$10^{-4}$, $10^{-5}$ from bottom to top. \vspace*{-.1in}}
\label{sommerm1}
\end{figure}

\begin{figure}[h]
\vspace*{-.01in} \centerline{
\includegraphics[width=3.3in,angle=0]{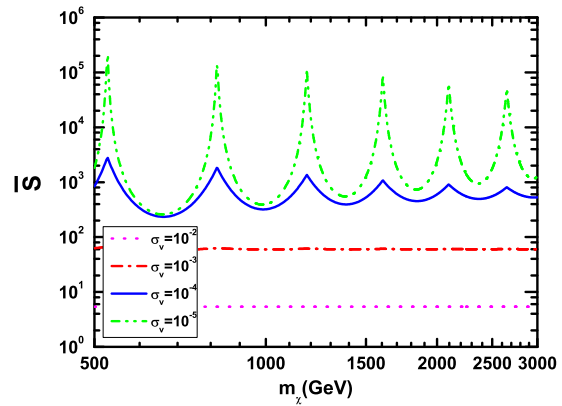}}
\vspace*{-.03in} \caption{Same with Fig.\ref{sommerm1}, but for the
averaged Sommerfeld enhancement factor $\bar{S}$
 as a function of DM
mass. Four curves denote different DM velocity dispersion $\sigma
_v$
 as
$10^{-2}$,
 $10^{-3}$,$10^{-4}$, $10^{-5}$ from bottom to top. \vspace*{-.1in}}
\label{sommerm}
\end{figure}

\begin{figure}[h]
\vspace*{-.01in}
\centerline{\includegraphics[width=3.3in,angle=0]{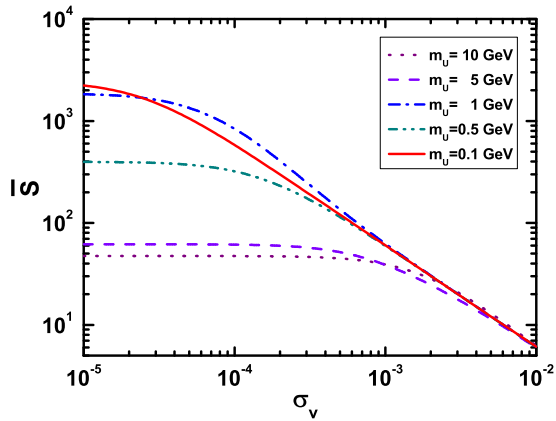}}
\vspace*{-.03in} \caption{The averaged Sommerfeld enhancement factor
$ \left\langle S \right\rangle$ as a function of DM velocity
dispersions $ \sigma _v$. Here we choose $m_\chi=1TeV$ and
$g_\chi=0.55$ ($ \alpha  = 2.41 \times 10^{ - 2}$ ). Five curves
denote different U-boson mass as $10 GeV$, $5GeV$, $0.5GeV$,
$0.1GeV$, $1GeV$, from bottom to top. \vspace*{-.1in}}
\label{sommerv}
\end{figure}

The numerical results of $ S$ and $ {\bar S}$
 are given in the Fig.\ref{sommerm1},
Fig.\ref{sommerm} and Fig.\ref{sommerv}. We can see that the
velocity distribution function $ f(v)$ has large possibility to have
velocity around $ O(\sigma _v )$, with the most probable velocity at
$ \sqrt 2 \sigma _v$. If the $ \sigma _v$ is very small, then $
f(v)$ behaves like some delta function around $ O(\sigma _v )$. In
this case, the $ {\bar S}$ has the same behavior with S. If the $
\sigma _v$ is quite large, the $ f(v)$ is a broad distribution
around $ O(\sigma _v )$. In this case, the $ {\bar S}$ is the
composition of different $ S(v)$. In Fig.\ref{sommerm1} and
Fig.\ref{sommerm}, the behavior of S and $ {\bar S}$ agree with the
above discussion.

To better understand the dependence of S on its four free
parameters, we discuss the behavior of S by simplifying the equation
under reasonable approximation. Since the light gauge boson is much
lighter than the DM, we can expand the Yukawa potential. The
Eq.\ref{radial} can then be written as,

\begin{equation}
\frac{1}{{m_\chi  }}\psi ''(r) + \frac{\alpha }{r}\psi (r) = ( -
m_\chi  \beta ^2  + \alpha m_{U}^{\;} )\psi (r) \label{radial2}.
\end{equation}

If $ \sqrt {\alpha m_{U}^{\;} /m_\chi  }  \ll \beta$, the behavior
of DM annihilation by exchange U-boson is similar with Coulomb
scattering. If $\alpha \ll \beta$ (and automatically $ \sqrt {\alpha
m_{U}^{\;} /m_\chi }  \ll \beta$, because $m_U \ll m_\chi$ in our
case), the enhancement can be negligible with $ S \simeq 1$. This is
the non-enhancement case. If $ \sqrt {\alpha m_{U}^{\;} /m_\chi }
\ll \beta \ll \alpha$, the $S$ is enhanced by $ 1/\beta$ with $ S
\simeq \pi \alpha /\beta$. This is the moderate enhancement case. In
the Fig.\ref{sommerm1}, the lines with $ \beta = 10^{ - 2} ,10^{ -
3}$ correspond to the moderate enhancement. In the
Fig.\ref{sommerv}, the lines with $ m_{U}^{\;}  = 10GeV$, $5GeV$,
$0.5GeV$ and $0.1GeV$ are also the cases of moderate enhancement.
Before the saturation, we can see $ {\bar S}$ grows with $ 1/\sigma
_v$ linearly. However, $ {\bar S}$ does not go to infinity because
it saturates at some small velocity dispersion. We can see if the
mass $ m_U^{\;}$ is small, the value of $ \sigma _v$ to reach the
saturate platform is also small.

If $ \sqrt {\alpha m_{U}^{\;} /m_\chi  }  \gg \beta$, the
Eq.\ref{radial2} has the similar form as the equation describing
hydrogen atom. The positiveness of the right hand side of the
equation points to the existence of bound states which can
significantly enhance the S \cite{Pieri:2009zi,MarchRussell:2008tu}.
The enhancement is finite due to the saturation in the low velocity
regime or finite width of the bound state. Close to the resonance, $
S \simeq \frac{{\alpha m_{U}^{\;} }}{{m_\chi  \beta ^2 }}$, which is
the resonance enhancement case. Recalling the energy level of
hydrogen atom $E_n$, the equation meets different resonances when we
vary the value of $ {m_\chi}$. Therefore the resonances appear
periodically in the Fig. \ref{sommerm1} and Fig. \ref{sommerm}. In
addition for $ \sqrt {\alpha m_{U}^{\;} /m_\chi } \gg \beta$, the
equation can neglect the term which contains $ \beta$. That is the
reason why the resonances locate at the same $ m_\chi$ for different
values of $ \beta$ or $ \sigma _v$ in the Fig. \ref{sommerm1} and
Fig. \ref{sommerm}. In the Fig. \ref{sommerv}, the line with $
m_{U}^{\;}  = 1GeV$ shows that it is close to a resonance when $
\sigma _v$ is low enough which makes it different from other four
moderate enhancement cases.

From Fig.\ref{sommerm1}, Fig.\ref{sommerm} and Fig.\ref{sommerv}, we
can see that the Sommerfeld enhancement is around $O(10^2)$ for a
typical DM velocity dispersion of $10^{-3}$ in the halo today. For
the lower velocity, the enhancement will increase significantly.
Such large enhancement is required to explain PAMELA/ATIC results.

\section{Searching for U-boson via $
e^ +  e^ -   \to U\gamma$ process }

At the BESIII detector, the luminosity of $ e^ +  e^ -$ collision is
$10^{33} cm^{ - 2} s^{ - 1}$ at $ \sqrt s  = 3.097GeV$. In our
numerical simulations we choose $ e^ +  e^ -$ integrated luminosity
as $20 fb^{ - 1}$ which corresponds to data samples collected within
four years. Throughout the paper, we utilize the package CalcHEP
\cite{Pukhov:2004ca} to simulate signal and corresponding background
processes after appropriate modifications of the model file.

In the model we adopted here, the U-boson does not directly couple
with quarks, so the signals and backgrounds are mainly leptons and
photons. Since the $\sqrt s$ which we adopted in this paper is lower
than $ 2 m_\tau$, the U-boson at BESIII can not decay into two tau
leptons. Thus, we do not take into account the tau signals. In the
visible decay channel, the U-boson decays to electrons and muons. In
the invisible decay channel, the U-boson decays to corresponding
neutrinos.

\subsection{Invisible decay mode $U \to \nu\bar \nu$ }

The signal process is
\begin{equation}
e^ +  e^ -   \to U\gamma  \to \nu\bar \nu \gamma .
\end{equation}

The main SM backgrounds for the signal process are
\begin{equation}
 e^ +  e^ -   \to \nu \bar \nu\gamma
\end{equation}

\begin{figure}[h]
\vspace*{-.03in} \centering
\includegraphics[width=0.90\columnwidth]{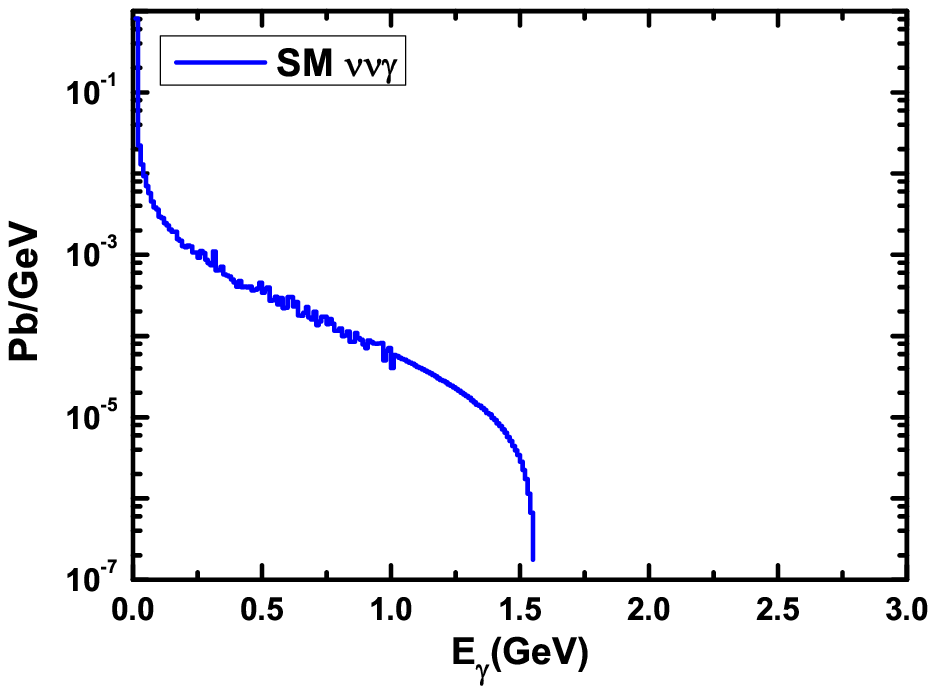}
\vspace*{-.03in} \caption{Photon energy distribution of SM
background for $e^ +  e^ -   \to \nu\bar \nu\gamma$ with $ \left|
{\cos \theta _\gamma  } \right| < 0.9$.
 \vspace*{-.1in}}
\label{invbkg}
\end{figure}

In Fig. \ref{invbkg} we show the photon energy distribution for
backgrounds with $ \left| {\cos \theta _\gamma  } \right| < 0.9$.
Here $ {\theta _\gamma  }$ corresponds to the angle among electron
beam line and photon. The background has the continuous photon
comparing with the mono-energetic photon from signal which has
energy $ E = (s - m_U^2 )/(2\sqrt s )$. Note that the background is
from higher-order contributions, i.e. $ O(\alpha G_F^2 s)$, compared
to the signal. At the BESIII, the energy resolution for electrons or
photons is about $ 2.3\% /\sqrt {E(GeV)}  \oplus 1\%$ with the
energy measurement range from $ 20MeV$ to $
2GeV$\cite{Asner:2008nq}. Therefore we impose the cuts as following,
\begin{equation}
{E_\gamma   > (s - m_U^2 )/(2\sqrt s ) - 0.2GeV}
\end{equation}
\begin{equation}
{\left| {\cos \theta _\gamma  } \right| < 0.9}.
\end{equation}

\begin{figure}[h]
\vspace*{-.03in} \centering
\includegraphics[width=0.90\columnwidth]{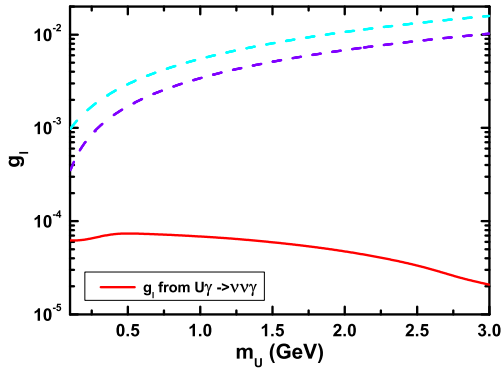}
\vspace*{-.03in} \caption{The solid line denotes $5\sigma$ lower
limit of $ {g_\ell}$ which can be detected by BESIII with $20fb^{ -
1}$ $ e^ + e^ -$ luminosity as a function of $ m_U$ for invisible
channel. The dash lines indicate the upper bounds from low-energy
experiments. The lower dash line shows the constraints from low
energy $\nu -e$ cross section; the upper one comes from the
measurement of $g_\mu-2$. We do not show the looser upper bound from
$g_e-2$ and $g_\tau-2$, since they are larger than $O(10^{-2})$
here.
 \vspace*{-.1in}}
\label{invsig}
\end{figure}

In Fig. \ref{invsig} we show the lower limit of $ {g_l}$ for
detecting U-boson as a function of $m_U$ with $ S/\sqrt{S+B} > 5$,
in which S and B represent the number of events for signal and
background respectively\footnote{In Ref.\cite{Zhu:2007zt}, there is
a typo for the definition of significance in invisible channel which
should be $ S/\sqrt{S+B}$. Our numerical results agree with the
result in Ref.\cite{Zhu:2007zt}. The reason for choosing $ S/\sqrt
{S+B}$ other than $ S/\sqrt B$ in the invisible channel is that the
background is extremely low. The number of background B is usually
below 1. However, in the visible channel, we choose $ S/\sqrt B$
because both signal and background have enough statistics.}. We also
give the possible constraints from the $g-2$ and low energy $\nu-e$
cross section.

\subsection{ Visible decay mode $U \to l\bar l $}

The U boson can decay into charged lepton pairs which has the signal
$ e^ + e^ -   \to U\gamma  \to l\bar l\gamma$. In the signal process
the $ m_{l\bar l}$ peaks around $ m_U$, and the SM background has
the smooth $ m_{l\bar l}$  except around $ \sqrt S$ and low energy
region due to t-channel contributions with soft photon and s-channel
contributions, respectively \cite{Zhu:2007zt}.

\begin{figure}[h]
\vspace*{-.03in} \centering
\includegraphics[width=0.90\columnwidth]{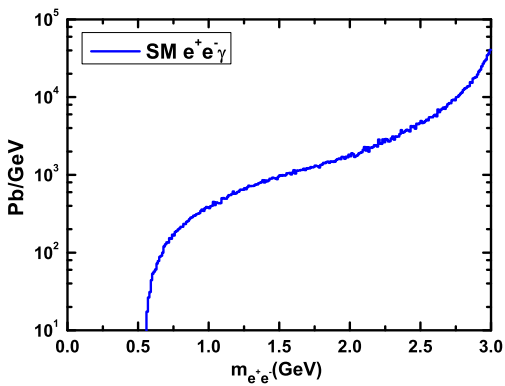}
\vspace*{-.03in} \caption{$ m_{e\bar e}$ distribution of the SM
background $ e^ +  e^ -   \to e^ +  e^ -  \gamma$ with the last two
cuts in Eqs.(\ref{visiblecut1})-(\ref{visiblecut3}).
 \vspace*{-.1in}}
\label{eeabkg}
\end{figure}

\begin{figure}[h]
\vspace*{-.03in} \centering
\includegraphics[width=0.90\columnwidth]{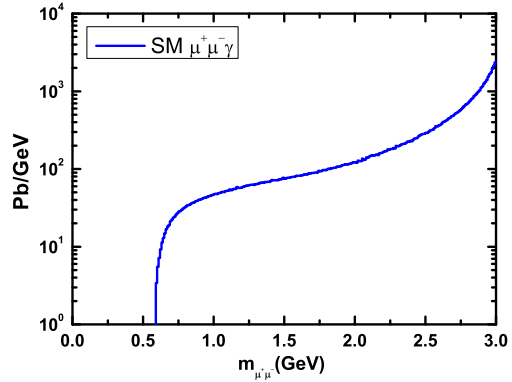}
\vspace*{-.03in} \caption{$ m_{\mu \bar \mu }$
 distribution of the SM
background $ e^ +  e^ -   \to \mu \bar \mu \gamma$
 with the last two cuts in Eqs.\ref{visiblecut1}-\ref{visiblecut3}.
 \vspace*{-.1in}}
\label{mmabkg}
\end{figure}
In Fig.\ref{eeabkg} and Fig.\ref{mmabkg}, we show the $ m_{e\bar e}$
and $ m_{\mu \bar \mu }$ distribution of the SM backgrounds. Since
the signal peaks around $ m_U$, the resolution of $ m_{l\bar l}$ is
important to suppress the backgrounds. To clearly separate the
electron and photon, the BESIII requires directions of two particle
has an open angle larger than $20^ \circ$ \cite{mao}. Thus we have
the following cut conditions,
\begin{equation}
\mid m_{l\bar l}  - m_U \mid < 1, 3 \;or\; 5 MeV,
\label{visiblecut1}
\end{equation}
\begin{equation}
\cos (\theta _i )< 0.9, \label{visiblecut2}
\end{equation}
\begin{equation}
\cos (\theta _{l\gamma } )< 0.94, \label{visiblecut3}
\end{equation}
where $ {\theta _i }$ ($ i = l,\bar l,\gamma$ ) corresponds to the
angles among initial electron beam line and final state particles
respectively. The $ {\theta _{l\gamma } }$ means the angle between
the lepton and photon in the final states. Many photons with energy
lower than $O(10)MeV$ come from the final state radiation of
$e^+e^-\rightarrow l^+l^-$. The direction of radiated $\gamma$ is
close to the direction of outgoing charge leptons. So it is obvious
to see that the Eq.(\ref{visiblecut3}) excludes most $l^+l^-\gamma$
events with low $m_{l^+l^-}$ in the Fig.\ref{eeabkg} and
Fig.\ref{mmabkg}. We give three kinds of ideal $ m_{l\bar l}$
resolution cuts which are $ {1MeV}$, $ {3MeV}$ and $ {5MeV}$. The
huge background has been suppressed at least two orders of magnitude
via cuts in Eqs.\ref{visiblecut1}-\ref{visiblecut3}.

\begin{figure}[h]
\vspace*{-.03in} \centering
\includegraphics[width=0.90\columnwidth]{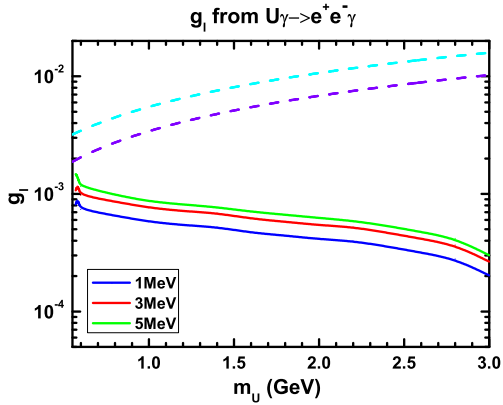}
\vspace*{-.03in} \caption{Same with Fig.\ref{invsig}, but for signal
channel $ e^ +  e^ -   \to U\gamma  \to e\bar e\gamma$. The solid
lines from bottom to top denote different cuts with $1MeV$, $3MeV$
and $5MeV$ in Eq.\ref{visiblecut1} respectively.
 \vspace*{-.1in}}
\label{ueesig}
\end{figure}

\begin{figure}[h]
\vspace*{-.03in} \centering
\includegraphics[width=0.90\columnwidth]{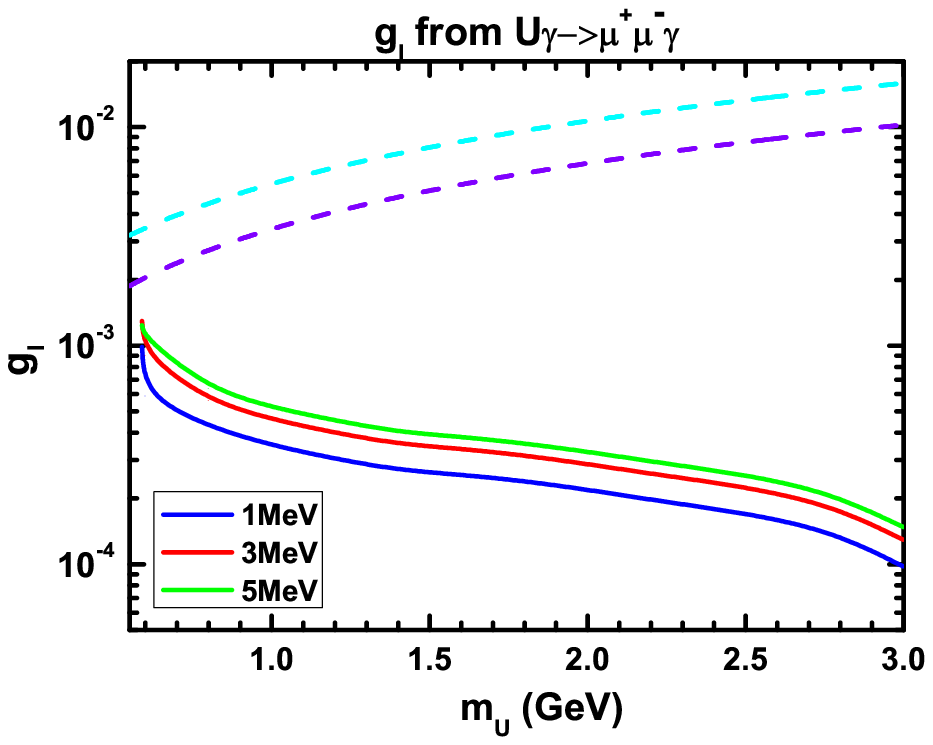}
\vspace*{-.03in} \caption{Same with Fig.\ref{ueesig}, but for signal
channel $ e^ + e^ - \to U\gamma  \to \mu \bar \mu \gamma$.
 \vspace*{-.1in}}
\label{ummsig}
\end{figure}

Fig.\ref{ueesig} and Fig.\ref{ummsig} show the lower limit of
${g_{\ell}}$ as a function of $ m_U$ with $ S/\sqrt B  > 5$ for
signal channel $ e^ +  e^ -   \to U\gamma  \to e\bar e\gamma$ and $
e^ + e^ - \to U\gamma  \to \mu \bar \mu \gamma$ respectively. The
conventions are the same with Fig.\ref{invsig}. We can see that the
${g_{l}}$ can reach  around $ 10^{ - 3}  \sim 10^{ - 4}$
 which is similar for the two visible channels. Note that
the invisible channel can reach the $ 10^{ - 4}  \sim 10^{ - 5}$
region because of the lower SM background.

\section{Conclusions and discussions}

In this paper we investigated one unified DM picture which can
account for the recent PAMELA/ATIC observations while still
consistent with other measurements, in a model with an extra
$U(1)_{L_i-L_j}$ gauge group with $L_i$ the $i-$th generation lepton
number. In order to obtain the boost factor (BF) via the so-called
Sommerfeld enhancement, the light new gauge boson $O(GeV)$ is
required, though the DM is around $O(TeV)$. After showing that the
required BF can be easily realized, we simulated the signal and
background of U-boson at BESIII detector. Our studies showed that it
is possible to detect such light leptophilic U-boson at the BESIII
via the process of $e^+e^-\rightarrow U\gamma$, followed by
$U\rightarrow e^+e^-$, $U\rightarrow \mu^+\mu^-$ and $U\rightarrow
\nu\overline{\nu}$.  All the U-boson decay modes can be utilized to
search for U-boson. For the U-boson invisible decay mode
$U\rightarrow \nu\overline{\nu}$, the BESIII can measure the
coupling of the extra $ U(1)$ down to $ O(10^{ - 4} ) \sim O(10^{ -
5} )$ with $5\sigma$ significance. For the charged lepton decay
modes, the  $5\sigma$ detecting limit can reach $10^{-4} \sim
10^{-3}$ for U-boson mass $m_U=0.5 GeV\sim 3GeV$. In fact, the
U-boson search at BESIII can also be carried out at $\sqrt{s}=2\sim5
GeV$. By the scanning of $\sqrt{s}$, it is even possible to detect
the U-boson by the $e^+e^-$ and/or $\mu^+\mu^-$ resonances.

Besides the low energy collider search for light leptophilic
U-boson, we would like to mention the interesting features in the DM
indirect and direct detection experiments. First, If the $U(1)$
group is gauged under $U(1)_{L_{\mu}-L_{\tau}}$, the final positron
spectrum from DM annihilation does not fit ATIC results very well.
It requires heavier DM than $ U(1)_{L_e  - L_\mu }$ and $ U(1)_{L_e
- L_\tau }$ cases, because the initial energy spectrum of positron
from $\mu$ or $\tau$ are quite soft. Second, because $U-\nu$
couplings equal to $U-l^\pm$, DM annihilations will produce high
energy neutrinos with energy of $m_\chi/2$. It is possible to detect
such neutrino flux in the next generation of neutrino telescopes
such as IceCube, Antares, $etc$ \cite{Liu:2008ci}(Moreover, the
$U(1)_{L_{i}-L_{j}}$ will induce the interaction between the
high-energy neutrinos and the background neutrinos. Measuring
high-energy cosmic neutrino flux spectrum at neutrino telescopes may
find an absorption feature due to the new $U(1)$ interaction
\cite{Hooper:2007jr}.) On the other hand, the
Super-Kamiokande(Super-K) data of neutrinos from the Galaxy
Center(GC) \cite{Desai:2004pq} can be used to constrain the model.
If the U-boson have decay channel to electron/positron, we only need
a boost factor of a few hundreds to explain PAMELA/ATIC results,
since the positron spectrum from such decay channel is quite hard
\cite{Cholis:2008wq}. Fortunately, such boost factor does not
violate the Super-K limit, especially for DM profile which is smooth
in the GC \cite{Liu:2008ci,Yuksel:2007ac,Hisano:2008ah}. Third, the
$\chi-e$ interaction may induce visible leptonic recoils far larger
than nuclear recoils at the DM direct detection experiments, because
the DM only directly couples to leptophilic U-boson. This feature
may be used to explain the DAMA modulation signal
\cite{Bernabei:2007gr,Fox:2008kb}, but it still faces some problems
\cite{Cui:2009xq}.

Recently, Fermi\cite{Abdo:2009zk} and
H.E.S.S.\cite{Aharonian:2009ah} give their results on the electron
and positron flux. The sharp ATIC "bump" at $300\sim800GeV$ are not
reported. For annihilating DM, the $ e^ +  e^ -$ channel is not
eagerly needed since there is no peak. The $ \mu ^ +  \mu ^ -$ and $
\tau ^ + \tau ^ -$ channels are needed to fit the Fermi data. In
annihilation scenario, the light new mediating particle is usually
needed to provide the Sommerfeld enhancement. If it is an extra U(1)
gauge boson, it is difficult to avoid the decay to $ e^ + e^ -$ in
the scenarios where the new gauge boson couples to leptons via
kinetic mixing to photon. But it can avoid $ e^ + e^ -$ if this
extra U(1) assigns charge directly on leptons, like $ U(1)_{L_\mu -
L_\tau }$ discussed in this paper (unfortunately, such new gauge
boson in the model with extra $ U(1)_{L_\mu - L_\tau }$ would not
easily be produced in the low-energy $ e^ + e^ -$ colliders).
Interestingly, if it is a scalar boson which has mixing with Higgs,
it naturally avoids $ e^ + e^ -$ since the couplings with leptons
are proportional to lepton mass\cite{Kohri:2009bd,Bergstrom:2009fa}.
It should be mentioned that $ \mu ^ +  \mu ^ -$ and $ \tau ^ + \tau
^ -$ channels in annihilation scenario usually receive stringent
limits from gamma and neutrino observations, while decay scenario
can cleanly compatible with these
observations\cite{Bergstrom:2009fa,Meade:2009iu}. For decaying DM,
the $ \mu$ and $ \tau$ leptons are also needed to interpret Fermi.
This usually relies on some special requirements on the Yukawa
coupling coefficients\cite{Shirai:2009fq}.

\section{ Acknowledgements}

This work was supported in part by the Natural Sciences Foundation
of China (Nos. 10775001, 10635030).


\begin{thebibliography}{99}
\bibitem{Adriani:2008zq}
  O.~Adriani {\it et al.},
  arXiv:0810.4994 [astro-ph];
  O.~Adriani {\it et al.},
  arXiv:0810.4995 [astro-ph].

\bibitem{:2008zz}
  J.~Chang {\it et al.},
  Nature {\bf 456}, 362 (2008).


\bibitem{Hooper:2008kv}
  D.~Hooper, A.~Stebbins and K.~M.~Zurek,
  arXiv:0812.3202 [hep-ph].

\bibitem{Grajek:2008jb}
  P.~Grajek, G.~Kane, D.~J.~Phalen, A.~Pierce and S.~Watson,
  arXiv:0807.1508 [hep-ph].

\bibitem{ArkaniHamed:2008qn}
  N.~Arkani-Hamed, D.~P.~Finkbeiner, T.~Slatyer and N.~Weiner,
  arXiv:0810.0713 [hep-ph].

\bibitem{Pospelov:2008jd}
  M.~Pospelov and A.~Ritz,
  arXiv:0810.1502 [hep-ph].

\bibitem{sommerfeld}
  H.~Baer, K.~m.~Cheung and J.~F.~Gunion,
  Phys.\ Rev.\  D {\bf 59}, 075002 (1999)
  [arXiv:hep-ph/9806361];
  J.~Hisano, S.~Matsumoto, M.~M.~Nojiri and O.~Saito,
  Phys.\ Rev.\  D {\bf 71}, 063528 (2005)
  [arXiv:hep-ph/0412403];
  J.~Hisano, S.~Matsumoto, M.~Nagai, O.~Saito and M.~Senami,
  Phys.\ Lett.\  B {\bf 646}, 34 (2007)
  [arXiv:hep-ph/0610249];
  M.~Cirelli, A.~Strumia and M.~Tamburini,
  Nucl.\ Phys.\  B {\bf 787}, 152 (2007)
  [arXiv:0706.4071 [hep-ph]];
  J.~March-Russell, S.~M.~West, D.~Cumberbatch and D.~Hooper,
  JHEP {\bf 0807}, 058 (2008)
  [arXiv:0801.3440 [hep-ph]];
  M.~Cirelli, M.~Kadastik, M.~Raidal and A.~Strumia,
  arXiv:0809.2409 [hep-ph].

\bibitem{Chun:2008by}
  E.~J.~Chun and J.~C.~Park,
  JCAP {\bf 0902}, 026 (2009)
  [arXiv:0812.0308 [hep-ph]].

\bibitem{Baumgart:2009tn}
  M.~Baumgart, C.~Cheung, J.~T.~Ruderman, L.~T.~Wang and I.~Yavin,
  JHEP {\bf 0904}, 014 (2009)
  [arXiv:0901.0283 [hep-ph]].

\bibitem{Fox:2008kb}
  P.~J.~Fox and E.~Poppitz,
  arXiv:0811.0399 [hep-ph].

\bibitem{Baek:2008nz}
  S.~Baek and P.~Ko,
  arXiv:0811.1646 [hep-ph].

\bibitem{He:1990pn}
  X.~G.~He, G.~C.~Joshi, H.~Lew and R.~R.~Volkas,
  Phys.\ Rev.\  D {\bf 43} (1991) 22;
  X.~G.~He, G.~C.~Joshi, H.~Lew and R.~R.~Volkas,
  Phys.\ Rev.\  D {\bf 44} (1991) 2118;
  R.~Foot, X.~G.~He, H.~Lew and R.~R.~Volkas,
  Phys.\ Rev.\  D {\bf 50} (1994) 4571
  [arXiv:hep-ph/9401250];
S.~Baek, N.~G.~Deshpande, X.~G.~He and P.~Ko,
  Phys.\ Rev.\  D {\bf 64} (2001) 055006
  [arXiv:hep-ph/0104141];

\bibitem{Bi:2009uj}
  X.~J.~Bi, X.~G.~He and Q.~Yuan,
  arXiv:0903.0122 [hep-ph].

\bibitem{Boehm:2003bt}
  C.~Boehm, D.~Hooper, J.~Silk, M.~Casse and J.~Paul,
  Phys.\ Rev.\ Lett.\  {\bf 92}, 101301 (2004)
  [arXiv:astro-ph/0309686].

\bibitem{Fayet:2007ua}
  P.~Fayet,
  Phys.\ Rev.\  D {\bf 75}, 115017 (2007)
  [arXiv:hep-ph/0702176].

\bibitem{EarlyUBoson}
P.~Fayet,
  Phys.\ Lett.\ B {\bf 95}, 285 (1980);
  P.~Fayet,
  Nucl.\ Phys.\ B {\bf 187}, 184 (1981).

\bibitem{Gninenko:1998pm}
  S.~N.~Gninenko and N.~V.~Krasnikov,
  Phys.\ Lett.\  B {\bf 427}, 307 (1998)
  [arXiv:hep-ph/9802375];
  S.~N.~Gninenko and N.~V.~Krasnikov,
  Phys.\ Lett.\  B {\bf 513}, 119 (2001)
  [arXiv:hep-ph/0102222].

\bibitem{Bouchiat:2004sp}
  C.~Bouchiat and P.~Fayet,
  Phys.\ Lett.\  B {\bf 608}, 87 (2005)
  [arXiv:hep-ph/0410260];
  P.~Fayet,
  arXiv:hep-ph/0607094;
  P.~Fayet,
  Phys.\ Rev.\ D {\bf 74}, 054034 (2006)
  [arXiv:hep-ph/0607318].

\bibitem{Borodatchenkova:2005ct}
  N.~Borodatchenkova, D.~Choudhury and M.~Drees,
  Phys.\ Rev.\ Lett.\  {\bf 96}, 141802 (2006)
  [arXiv:hep-ph/0510147].

\bibitem{Zhu:2007zt}
  S.~h.~Zhu,
  Phys.\ Rev.\  D {\bf 75}, 115004 (2007)
  [arXiv:hep-ph/0701001].

\bibitem{Chen:2007uv}
  C.~H.~Chen, C.~Q.~Geng and C.~W.~Kao,
  Phys.\ Lett.\  B {\bf 663}, 400 (2008)
  [arXiv:0708.0937 [hep-ph]].

\bibitem{Boehm:2003hm}
  C.~Boehm and P.~Fayet,
  Nucl.\ Phys.\ B {\bf 683}, 219 (2004)
  [arXiv:hep-ph/0305261].

\bibitem{ArkaniHamed:2008qp}
  N.~Arkani-Hamed and N.~Weiner,
  JHEP {\bf 0812}, 104 (2008)
  [arXiv:0810.0714 [hep-ph]].

\bibitem{Pospelov:2008zw}
  M.~Pospelov,
  arXiv:0811.1030 [hep-ph].

\bibitem{Batell:2009yf}
  B.~Batell, M.~Pospelov and A.~Ritz,
  arXiv:0903.0363 [hep-ph].

\bibitem{Essig:2009nc}
  R.~Essig, P.~Schuster and N.~Toro,
  arXiv:0903.3941 [hep-ph].

\bibitem{Reece:2009un}
  M.~Reece and L.~T.~Wang,
  arXiv:0904.1743 [hep-ph].

\bibitem{Hanneke:2008tm}
  D.~Hanneke, S.~Fogwell and G.~Gabrielse,
  Phys.\ Rev.\ Lett.\  {\bf 100}, 120801 (2008)
  [arXiv:0801.1134 [physics.atom-ph]].

\bibitem{Bennett:2006fi}
  G.~W.~Bennett {\it et al.}  [Muon G-2 Collaboration],
  Phys.\ Rev.\  D {\bf 73}, 072003 (2006)
  [arXiv:hep-ex/0602035].

\bibitem{Abdallah:2003xd}
  J.~Abdallah {\it et al.}  [DELPHI Collaboration],
  Eur.\ Phys.\ J.\  C {\bf 35}, 159 (2004)
  [arXiv:hep-ex/0406010].

\bibitem{Allen:1992qe}
  R.~C.~Allen {\it et al.},
  Phys.\ Rev.\  D {\bf 47}, 11 (1993).
  L.~B.~Auerbach {\it et al.}  [LSND Collaboration],
  Phys.\ Rev.\  D {\bf 63}, 112001 (2001)
  [arXiv:hep-ex/0101039].

\bibitem{Jungman:1995df}
  G.~Jungman, M.~Kamionkowski and K.~Griest,
  Phys.\ Rept.\  {\bf 267}, 195 (1996).
  [arXiv:hep-ph/9506380].

\bibitem{Iengo:2009ni}
  R.~Iengo,
  JHEP {\bf 0905}, 024 (2009)
  [arXiv:0902.0688 [hep-ph]].

\bibitem{Pieri:2009zi}
  L.~Pieri, M.~Lattanzi and J.~Silk,
  arXiv:0902.4330 [astro-ph.HE].

\bibitem{Bovy:2009zs}
  J.~Bovy,
  arXiv:0903.0413 [astro-ph.HE].

\bibitem{MarchRussell:2008tu}
  J.~D.~March-Russell and S.~M.~West,
  arXiv:0812.0559 [astro-ph].

\bibitem{Pukhov:2004ca}
  A.~Pukhov,
  arXiv:hep-ph/0412191.

\bibitem{Asner:2008nq}
  D.~M.~Asner {\it et al.},
  arXiv:0809.1869 [hep-ex].

\bibitem{mao}
  Private communication with Ya-jun Mao.

\bibitem{Liu:2008ci}
  J.~Liu, P.~f.~Yin and S.~h.~Zhu,
  arXiv:0812.0964 [astro-ph].

\bibitem{Hooper:2007jr}
  D.~Hooper,
  Phys.\ Rev.\  D {\bf 75}, 123001 (2007)
  [arXiv:hep-ph/0701194].

\bibitem{Desai:2004pq}
  S.~Desai {\it et al.}  [Super-Kamiokande Collaboration],
  Phys.\ Rev.\  D {\bf 70}, 083523 (2004)
  [Erratum-ibid.\  D {\bf 70}, 109901 (2004)]
  [arXiv:hep-ex/0404025].

\bibitem{Cholis:2008wq}
  I.~Cholis, G.~Dobler, D.~P.~Finkbeiner, L.~Goodenough and N.~Weiner,
  arXiv:0811.3641 [astro-ph].

\bibitem{Yuksel:2007ac}
  H.~Yuksel, S.~Horiuchi, J.~F.~Beacom and S.~Ando,
  Phys.\ Rev.\  D {\bf 76}, 123506 (2007)
  [arXiv:0707.0196 [astro-ph]].

\bibitem{Hisano:2008ah}
  J.~Hisano, M.~Kawasaki, K.~Kohri and K.~Nakayama,
  arXiv:0812.0219 [hep-ph].

\bibitem{Bernabei:2007gr}
  R.~Bernabei {\it et al.},
  Phys.\ Rev.\  D {\bf 77}, 023506 (2008)
  [arXiv:0712.0562 [astro-ph]].

\bibitem{Cui:2009xq}
  Y.~Cui, D.~E.~Morrissey, D.~Poland and L.~Randall,
  arXiv:0901.0557 [hep-ph].

\bibitem{Abdo:2009zk}
  A.~A.~Abdo {\it et al.}  [The Fermi LAT Collaboration],
  arXiv:0905.0025 [astro-ph.HE].

\bibitem{Aharonian:2009ah}
  H.~E.~S.~Aharonian,
  arXiv:0905.0105 [astro-ph.HE].

\bibitem{Bergstrom:2009fa}
  L.~Bergstrom, J.~Edsjo and G.~Zaharijas,
  arXiv:0905.0333 [astro-ph.HE].

\bibitem{Kohri:2009bd}
  K.~Kohri, J.~McDonald and N.~Sahu,
  arXiv:0905.1312 [hep-ph].

\bibitem{Meade:2009iu}
  P.~Meade, M.~Papucci, A.~Strumia and T.~Volansky,
  arXiv:0905.0480 [hep-ph].

\bibitem{Shirai:2009fq}
  S.~Shirai, F.~Takahashi and T.~T.~Yanagida,
  arXiv:0905.0388 [hep-ph].

\end{thebibliography}
\end{document}